\documentstyle[12pt,aasms4,eqsecnum]{article}
\lefthead{Miralles and Van Riper}
\righthead{Fermi-Dirac Integrals and Derivatives}
\begin{document}

\title{Accurate Evaluation of\\ Fermi-Dirac Integrals and
       Their Derivatives\\ for Arbitrary Degeneracy and Relativity
       (astro-ph/9509124)}

\author{Juan Antonio Miralles\altaffilmark{1}}
\affil{Departamento de Astronom\'{\i}a y Astrof\'{\i}sica,
        Universitat de Val\`encia,\\
        Dr. Moliner 50, 46100--Burjassot (Valencia), SPAIN}

\and

\author{Kenneth A. Van Riper}
\affil{Transport Methods Group, Los Alamos National Laboratory\\
        MS B226,  Los Alamos, NM  87545, USA}

\altaffiltext{1}{IGPP, Los Alamos National Laboratory}
%
%
\def\fdarg{(\eta,\beta)}
\def\shillfrac#1#2{#1/#2}
\def\fone{F_{\shillfrac{1}{2}}\fdarg}
\def\fthree{F_{\shillfrac{3}{2}}\fdarg}
\def\ffive{F_{\shillfrac{5}{2}}\fdarg}
\def\fsone{f_{\shillfrac{1}{2}}}
\def\fsthree{f_{\shillfrac{3}{2}}}
\def\fsfive{f_{\shillfrac{5}{2}}}
\def\dfoneeta{{\partial F_{\shillfrac{1}{2}}\fdarg\over \partial \eta}}
\def\dfthreeeta{{\partial F_{\shillfrac{3}{2}}\fdarg\over \partial \eta}}
\def\dffiveeta{{\partial F_{\shillfrac{5}{2}}\fdarg\over \partial \eta}}
\def\dfonebeta{{\partial F_{\shillfrac{1}{2}}\fdarg\over \partial \beta}}
\def\dfthreebeta{{\partial F_{\shillfrac{3}{2}}\fdarg\over \partial \beta}}
\def\dffivebeta{{\partial F_{\shillfrac{5}{2}}\fdarg\over \partial \beta}}
\def\foneth{F_{\shillfrac{1}{2}}^{\hbox{th}}\fdarg}
\def\fthreeth{F_{\shillfrac{3}{2}}^{\hbox{th}}\fdarg}
\def\ffiveth{F_{\shillfrac{5}{2}}^{\hbox{th}}\fdarg}
\def\dfonebetath{{\partial F_{\shillfrac{1}{2}}^{\hbox{th}}\fdarg\over \partial
\beta}}
\def\dfthreebetath{{\partial F_{\shillfrac{3}{2}}^{\hbox{th}}\fdarg\over
\partial \beta}}
\def\dffivebetath{{\partial F_{\shillfrac{5}{2}}^{\hbox{th}}\fdarg\over
\partial \beta}}
\def\dfoneetath{{\partial F_{\shillfrac{1}{2}}^{\hbox{th}}\fdarg\over \partial
\eta}}
\def\dfthreeetath{{\partial F_{\shillfrac{3}{2}}^{\hbox{th}}\fdarg\over
\partial \eta}}
\def\dffiveetath{{\partial F_{\shillfrac{5}{2}}^{\hbox{th}}\fdarg\over \partial
\eta}}
\def\ghalf{G_{\shillfrac{1}{2}}(\eta)}
\def\gone{G_1(\eta)}
\def\gthreeh{G_{\shillfrac{3}{2}}(\eta)}
\def\gtwo{G_2(\eta)}
\def\gfiveh{G_{\shillfrac{5}{2}}(\eta)}
\def\gthree{G_3(\eta)}
\def\ghalfeta{{\partial G_{\shillfrac{1}{2}}(\eta)\over \partial \eta}}
\def\goneeta{{\partial G_1(\eta)\over \partial \eta}}
\def\gthreeheta{{\partial G_{\shillfrac{3}{2}}(\eta)\over \partial \eta}}
\def\gtwoeta{{\partial G_2(\eta)\over \partial \eta}}
\def\gfiveheta{{\partial G_{\shillfrac{5}{2}}(\eta)\over \partial \eta}}
\def\gthreeeta{{\partial G_3(\eta)\over \partial \eta}}
\def\gfiveeta{{\partial G_{\shillfrac{5}{2}}(\eta)\over \partial \eta}}
\def\ki{K_I}
\def\kii{K_{II}}
\def\emass{m}
\def\cboltz{k_{\rm B}}
\def\bmass{m_{\rm o}}
\def\hplanck{h}
\def\clight{c}
\def\sqtoi{S}
\def\cnste{{1\over\sqrt{2}}}
\def\cnstf{{\pi^2\over6\sqrt{2}}}
\def\cnstg{{\pi^2\over6}}
%
%
\begin{abstract}
The equation of state of an ideal Fermi gas is expressed in terms of
Fermi-Dirac integrals.
We give formulae for evaluation the Fermi-Dirac integrals
of orders $\slantfrac{1}{2}$, $\slantfrac{3}{2}$, and $\slantfrac{5}{2}$
and their derivatives in various
limits of non- and extreme degeneracy and relativity.  We provide
tables and a Fortran subroutine for numerical evaluation of the integrals
and derivatives when a limit does not apply.  The functions can be
evaluated to better than 1\%\ accuracy for any temperature and density
using these methods.
\end{abstract}


\section{Introduction}
The electron constituent of neutron star envelopes exists in various
stages of degeneracy, from a classical gas on the surface to an
extremely degenerate Fermi gas in the interior.  Convective regimes
may exist within the envelope (\markcite{urpin} Urpin 1981).
Whether convection does occur depends on the magnitude of
the temperature gradient compared to the adiabatic gradient, which is
most conveniently calculated from the adiabatic index
\begin{equation}\openup2\jot
   \Gamma_2 = {1 \over  1 - \strut{1 / G } }
\end{equation}
where
\begin{equation}\openup2\jot
    G = {T\over P} {\partial P\over\partial T}\bigg\vert_\rho
          + {\rho\over P} {\partial P\over\partial\rho}\bigg\vert_T
           \left[\left({\partial E\over\partial\rho}\bigg\vert_T
         \right)\Bigg/\left( {\partial
P\over\partial\rho}\bigg\vert_T\right)\right]
%
\end{equation}
An estimate of the role of convection requires accurate values for the
temperature $T$ and density $\rho$ derivatives of the Pressure $P$ and
energy density $E$.
The envelope constituents are electrons, ions, and radiation.
The total equation of state, which we treat elsewhere, also contains
contributions from the Coulomb correction to the ions and takes into
account the changes in binding energy and electron density due to
the variation of ionization level with $T$ and $\rho$.

The electrons make a significant contribution to the equation of state (EOS)
in the nondegenerate regime, and dominate when they are degenerate.
We treat the electrons as an ideal Fermi gas of arbitrary degeneracy and
relativity.
We have previously evaluated the nonderivative electron EOS by a scheme which
interpolates in a two-dimensional table of Fermi-Dirac integrals
(eq. [\ref{eq:fermidiraci}] below) or uses
asymptotic formulae in degenerate, nondegenerate, relativistic, and
nonrelativistic
limits, where applicable.  Those limiting formulae are described in
\markcite{bludman} Bludman \&\ Van Riper (1977).

The evaluation scheme for the direct function is not sufficiently accurate
for the calculation of the EOS derivatives by direct numerical differences.
Such a scheme becomes ever more hopeless as the degeneracy increases.
The fundamental problem is obtaining temperature derivatives of quantities
which have a vanishingly small temperature dependence.  This problem also
presented itself in our derivation of asymptotic formulae for the degenerate
$T$
derivatives.  Since \markcite{urpin} Urpin (1981) had suggested
the possibility of convection in the
degenerate regime, we were particularly interested in obtaining an
accurate adiabatic index there.

Our method of evaluating the derivatives of the EOS relies on the derivatives
of
the Fermi-Dirac integrals, which are found by a scheme exactly analogous to
our method for the integrals themselves.  We prepare a table of the derivatives
by accurate numerical integration for intermediate degeneracy and relativity
and derived formulae (sometimes relying on other tabulated functions) in the
various limits.  The interpolation can be made with either a fast
second order method or with more accurate polynomial schemes.

We present those formulae here.  We will also publish our tables of the
Fermi-Dirac functions, their derivatives, modified Bessel functions of the
first and second kind, and some Fermi integrals on the Astrophysical Journal
CDROM.  We will also include FORTRAN subroutines in which our method
is implemented.
This paper serves as documentation for both tables and subroutines.
As such, we will give a brief review of the casting of the EOS in terms of
the Fermi-Dirac functions and will also show the limiting formulae for the
functions along with the corresponding formulae for their derivatives.

The monikers ``Fermi-Dirac'' and ``Fermi'' receive no
consistent usage in the literature.  We reserve the term
{\it Fermi-Dirac} integral (or function) for the bivariate (temperature and
degeneracy parameter $\eta$) functions defined in equation
(\ref{eq:fermidiraci}).  In the relativistic and nonrelativistic limits,
these functions reduce to expressions involving integrals
(eq. [\ref{eq:fermii}] below) which depend on  $\eta$ alone; we
refer to these latter as  integrals as the {\it Fermi} functions.

Numerous studies of the Fermi-Dirac integrals and methods for their rapid
evaluation exist in the literature (though few of these explicitly treat the
derivatives).   An excellent general reference is chapter 24 of the book of
\markcite{cox} Cox \&\ Giuli (1968),
which describes the general theory, formulae for many limiting
cases, and tabulations of the Fermi-Dirac integrals (based on unpublished work
by Terry W. Edwards).   The literature
contains a number of schemes for calculating one or more of the integrals.
\markcite{kippenhahn} Kippenhahn \&\ Thomas  (1964;
see also \markcite{kippenhahnetal} Kippenhahn, et al. 1967) give a power
series expansion for the
evaluation of the thermodynamic quantities $n$, $P$, and $E$  which is valid
for non- and mildly degenerate gases.
\markcite{divine} Divine (1965) gives a method, based on third order rational
function approximations to Fermi integrals  of orders
 $\slantfrac{1}{2}$, $\slantfrac{3}{2}$, 2, $\slantfrac{5}{2}$, and 3,
which gives $n$, $P$, and $E$ to a stated accuracy of 0.3\% for arbitrary
degrees of degeneracy and relativity.
\markcite{guess} Guess (1966) considers an set of functions ($Q_n$)
equivalent to the
Fermi-Dirac functions, and gives series expansions for high and low
temperatures
and degeneracies, accompanied by a tables for the central regime where none
of those limits apply.
\markcite{tooper} Tooper (1969), considering relativistic gases for
arbitrary degeneracy,
derives series expansions in the non- and extremely-degenerate limits
and discusses methods for numerical integration in the intermediate regime.
\markcite{beaudet}
Beaudet \&\ Tassoul (1971) give simple formulae with which $n$, $P$, and $E$
can
be evaluated to an accuracy of several percent for the relativistic and/or
degenerate regimes.  \markcite{bludman}
Bludman \&\ Van Riper (1977) give simple formulae, accurate to 0.5\%
for the semi-degenerate, relativistic and nonrelativistic  regimes.
\markcite{nadyozhin} Nadyozhin (1974) and \markcite{blinnikov}
Blinnikov \&\ Rudzskii (1988) consider the limiting cases
of extreme relativity and give a number of series expansions.
\markcite{eff} Eggleton, et al. (1973) derived
fitting formulae for
the $n$,  $P$, and $E$ of an ideal electron gas for a range of $T$ and $\rho$
covering the
regime of arbitrary degeneracy and relativity.
Evaluation of the thermodynamic quantities with 5$^{\hbox{th}}$ order
formulae (their
Table 4) agree with values based on our main table  numerical integrations to
0.04\%.  An extension of a 4$^{\hbox{th}}$ order method of Eggleton, et al.
given by \markcite{pols}
Pols et al. (1995, Appendix A), while thermodynamically consistent,
agrees with our integrations to 0.3\%.

There is a considerable literature dealing with the Fermi integrals.
We do not mention any of this work here, except to take note of
\markcite{antia}Antia's (1993)
rational function approximations for Fermi
integrals of several half-integral orders with stated maximum relative error of
$10^{-12}$.

In the next \S\ we give the formulae for several thermodynamic quantities
of an ideal Fermi gas in terms of the Fermi-Dirac integrals.
Asymptotic formulae in the degenerate limit are presented in \S\ 3, including
special treatment necessary for the temperature derivatives of the pressure
and energy density.   The treatments in other asymptotic regions are discussed
in
\S 4.  Section 5 covers the numerical details,
including the integration and interpolation methods and the accuracy
and efficiency of the scheme
for various interpolation orders.

\section{Thermodynamics of an Ideal Fermi Gas}
We will work throughout in terms of the dimensionless
degeneracy and temperature
      \begin{equation} \eta = {\mu\over \cboltz T}, \quad \hbox{and}\quad
         \beta = {\cboltz T\over mc^2}, \end{equation}
where $\mu$ is the chemical potential and $m$ is the mass of the fermion
(this theory is also applicable to ideal neutron and proton gases).
Constants such as the Boltzman constant $\cboltz$ have their usual
meaning throughout this paper.
The gas is degenerate (nondegenerate) for
 $\eta \gg 0$ ($\eta \ll 0$).
We will refer to relativity regimes based on the value
of $\beta$, with the gas being relativistic (nonrelativistic) for
$\beta \gg 1$ ($\beta\ll 1$).  The gas also becomes relativistic
at high density when the $\mu \gg \emass\clight^2$ or, equivalently, when
$\eta\beta \gg 1$.
In practice, we take the degenerate (nondegenerate) regime to be
$\eta \ge 70$ ($\eta \le -30$) and the relativistic (nonrelativistic)
regime to be $\beta \ge 10^4$ ($\beta \le 10^{-6}$).

The zero of energy for the particles is chosen so that the thermodynamic
potential is
\begin{equation}\Omega = -V\cboltz T \int {gd^3p\over h^3}
      \ln\left[1+\exp\left(\mu - \epsilon \right)\over\cboltz T\right],
\end{equation}
where  $p$ is the momentum, $g$ is the statistical weight, and
\begin{equation}
\epsilon = \sqrt{(\emass\clight^2)^2 + (p\clight)^2} - \emass\clight^2
\end{equation}
is the kinetic energy.
With the energy so defined, $\mu$ does not contain the rest mass.
(We do not consider antiparticles---positrons---in this work;
neutron star envelopes do not encounter the high $T$ and low $\rho$
where $e^+$ appear in significant numbers.)

The number density $n$, pressure $P$, and
energy density (per volume) $E$
of an ideal Fermi gas are
\begin{equation}
n = {8 \pi \sqrt{2} \emass^3 \clight^3 \over\hplanck^3}
 \beta^{3\over2} \left[\fone+\beta \fthree\right],
\end{equation}
\begin{equation}
P = {16 \pi \sqrt{2} \emass^4 \clight^5\over3\hplanck^3}
    \beta^{5\over2} \left[\fthree+{1\over2} \beta \ffive\right],
\end{equation}
and
\begin{equation}
E = {8 \pi \sqrt{2} \emass^4 \clight^5\over\hplanck^3}
      \beta^{5\over2} \left[\fthree+\beta \ffive\right],
\end{equation}
where the Fermi-Dirac integral of order $k$ is defined as
\begin{equation}
    F_k(\eta,\beta)=\int_{0}^{\infty}
      {x^k \left(1+{1\over2}\beta x \right)^{1\over2}\over{\exp(x-\eta)+1}}
\,dx.
             \label{eq:fermidiraci}
\end{equation}

The energy and pressure derivatives with respect to $T$ and $n$ are
\begingroup
\def\fdarg{}
\begin{equation} \openup2\jot
{\partial P\over\partial n}\bigg\vert_T = {2\over3} \emass \clight^2
         \beta \left(\dfthreeeta+{1\over2} \beta \dffiveeta
         \right)\Bigg/\left(
          \dfoneeta+\beta \dfthreeeta\right),
%
\end{equation}
\begin{equation} \openup2\jot
{\partial E\over\partial n}\bigg\vert_T = \emass \clight^2
        \beta \left(\dfthreeeta + \beta \dffiveeta
        \right)\Bigg/\left(
        \dfoneeta + \beta \dfthreeeta\right),
%
\end{equation}
\begin{eqnarray}
{\partial P\over\partial T}\bigg\vert_n = &
  { \displaystyle 16 \pi \sqrt{2} \emass^3 \clight^3
    \over \displaystyle 3\hplanck^3} \cboltz  \beta^{3\over2}
   \left\{\displaystyle{5\over2} \fthree + { \displaystyle\beta \dfthreebeta}
         + {7\over4} \beta \ffive
   + {1\over2} { \displaystyle\beta^2\dffivebeta}\right.
     \nonumber \\  \noalign{\medskip}& \left.
  - {\left( \displaystyle\dfthreeeta+{1\over2} \beta\dffiveeta\right)
   \left( \displaystyle {3\over2}\fone+\beta \dfonebeta + {5\over2} \beta
\fthree
         + \beta^2 \dfthreebeta\right)
   \over\left( \displaystyle\dfoneeta + \beta \dfthreeeta\right)}\right\},
   \label{eq:dpdt}
\end{eqnarray}
and
\begin{eqnarray}
{\partial E\over\partial T}\bigg\vert_n = &
  { \displaystyle8 \pi \sqrt{2} \emass^3 \clight^3
     \over \displaystyle\hplanck^3} \cboltz  \beta^{3\over2}
    \left\{\displaystyle{5\over2} \fthree + \beta { \displaystyle\dfthreebeta}
        + {7\over2} \beta \ffive + \beta^2 { \displaystyle\dffivebeta}
                                                    \right.\nonumber \\
     \noalign{\medskip}  &  \left.
   - {\left( \displaystyle\dfthreeeta+\beta \dffiveeta\right)
    \left( \displaystyle{3\over2} \fone
    + \beta \dfonebeta+{5\over2} \beta \fthree
    + \beta^2 \dfthreebeta\right)\over
    \left( \displaystyle\dfoneeta + \beta \dfthreeeta\right)}\right\},
   \label{eq:dedt}
\end{eqnarray}
\endgroup

    where the derivatives of the Fermi-Dirac integrals are
\begin{equation}
    {\partial F_k\over \partial \eta}(\eta,\beta) =
       \int_{0}^{\infty}{x^k \left(1+{1\over2}\beta x\right)^{1\over2}\over
         \left[\exp\left({x-\eta\over2}\right)
        +\exp\left({\eta-x\over2}\right)\right]^2}\,dx
\end{equation}
     and
\begin{equation}
    {\partial F_k\over \partial \beta}(\eta,\beta) =
       \int_{0}^{\infty}{x^{(k+1)} \left(1+{1\over2}\beta x\right)^{-{1\over2}}
                      \over4 \left[\exp(x-\eta)+1\right]}\,dx.
\end{equation}

\section{The Degenerate Limit}
\subsection{Temperature Derivatives}
The temperature dependence of $P$ and $E$ in the degenerate regime is
vanishingly small; obtaining accurate temperature derivatives is accordingly
problematic.  In particular,  the computer representation of
temperature-independent terms in (1) and (2) lacks sufficient resolution
to ensure cancellations which should occur.
We implement these cancellations analytically and use the resulting
{\it thermal} terms in the degenerate temperature derivatives.
When $\eta > 70$, the following are used
instead of (\ref{eq:dpdt}) and (\ref{eq:dedt}):
\begingroup
\def\fdarg{}
\begin{mathletters}
\begin{equation}
 {\partial P\over\partial T}\bigg\vert_{n,{\rm ED}}
   = {16 \pi \sqrt{2} \emass^3 \clight^3
      \over 3 \hplanck^3} \cboltz \beta^{3\over2}
      \left( T_1 - T_2 + T_3 - T_4\right)\Bigg/
        \left(\dfoneeta + \beta
      \dfthreeeta\right)
\end{equation}
where the terms in the numerator are
\begin{eqnarray}
T_1&=& \left( {5\over2} \fthree + \beta \dfthreebeta
	 + { 7\over4} \beta \ffive + {\beta^2\over2}  \dffivebeta \right)
	  \left(  \dfoneetath + \beta \dfthreeetath\right),
	 \nonumber \\
T_2&=&  \left( {3\over2} \fone + \beta \dfonebeta
         + {5\over2} \beta \fthree
	+ \beta^2 \dfthreebeta\right)
	\left(\dfthreeetath+{1\over2} \beta \dffiveetath \right),
         \nonumber \\
T_3&=&  \left(\dfoneeta + \beta \dfthreeeta - \dfoneetath -
	   \beta \dfthreeetath \right)
	   \left( {5\over2} \fthreeth + \beta \dfthreebetath
	   + {7\over4} \beta \ffiveth + {\beta^2\over2}  \dffivebetath \right),
	   \nonumber \\
T_4&=&  \left(\dfthreeeta+{1\over2} \beta \dffiveeta
	    - \dfthreeetath - {1\over2} \beta \dffiveetath \right)
         \left( {3\over2} \foneth + \beta \dfonebetath +
	   {5\over2} \beta \fthreeth + \beta^2 \dfthreebetath \right),
         \nonumber \\
 & &
\end{eqnarray}
\end{mathletters}
and
\begin{mathletters}
\begin{equation}
 {\partial E\over\partial T}\bigg\vert_{n,{\rm ED}}
    = {8 \pi \sqrt{2} \emass^3 \clight^3
       \over 3 \hplanck^3} \cboltz \beta^{3\over2}
	\left( T_5 - T_6 + T_7 - T_8\right)\Bigg/\left(\dfoneeta +
	\beta \dfthreeeta\right),
\end{equation}
where
\begin{eqnarray}
T_5&=& \left( {5\over2} \fthree + \beta \dfthreebeta
	 + {7\over2} \beta \ffive + \beta^2  \dffivebeta \right)
	  \left( \dfoneetath + \beta \dfthreeetath\right),
	 \nonumber \\
T_6&=&  \left( {3\over2} \fone + \beta \dfonebeta
       + {5\over2}  \beta \fthree
	+ \beta^2 \dfthreebeta\right)
	\left(\dfthreeetath+ \beta \dffiveetath \right),
         \nonumber \\
T_7&=&  \left(\dfoneeta + \beta \dfthreeeta - \dfoneetath -
	   \beta \dfthreeetath \right)
	   \left( {5\over2} \fthreeth
              + \beta \dfthreebetath
	   + {7\over2} \beta \ffiveth
          + \beta^2  \dffivebetath \right),
	   \nonumber \\
\noalign{\noindent and}
T_8&=&  \left(\dfthreeeta+ \beta \dffiveeta
	    - \dfthreeetath - \beta \dffiveetath \right)
	   \left( {3\over2} \foneth + \beta \dfonebetath +
	   {5\over2} \beta \fthreeth + \beta^2 \dfthreebetath \right).
         \nonumber \\
 & &
\end{eqnarray}
\end{mathletters}
\endgroup

\subsection{Degenerate Fermi-Dirac Functions}
The asymptotic degenerate formulae are given in terms of
\begin{equation}
          y =      \sqrt{A^2 - 1}
\end{equation}
where
\begin{equation}
          A =    1 + \eta \beta
\end{equation}
When the degenerate gas is relativistic, $\eta\beta \gg 1$ and
$y \approx \eta\beta \gg 1$.  Similarly, in the nonrelativistic limit,
$\eta\beta \ll 1$, $y \approx \sqrt{2\eta\beta} << 1$.  For most functions,
different formulae are used depending the value of $y$.

The following expressions for the
Fermi-Dirac functions and the thermal terms are only used in the
extreme degenerate limit ($\eta > 70$).  In that limit, the first few terms
in the degenerate expansions are  sufficient (additional terms in the expansion
may be found in \markcite{cox} Cox \&\ Giuli [1968, chapter 24]).
The expressions are
\begin{mathletters}
\begin{eqnarray}
\fone &\simeq&   \cnste \beta^{-{3\over2}} \fsone
          + \cnstf \eta^{-{1\over2}} {1+\eta\beta\over \sqrt{2+\eta\beta}} \\
           \noalign{\medskip}
\foneth &=& \cnstf \eta^{-{1\over2}} {1+\eta\beta\over \sqrt{2+\eta\beta}}
\end{eqnarray}
\end{mathletters}

\begin{mathletters}
\begin{eqnarray}
\fthree &\simeq&   \cnste \beta^{-{5\over2}} \fsthree
	   + \cnstf \eta^{1\over2} {3+2\eta\beta\over \sqrt{2+\eta\beta}}, \\
	 \noalign{\medskip}
\fthreeth &=& \cnstf \eta^{1\over2} {3+2\eta\beta\over \sqrt{2+\eta\beta}},
\end{eqnarray}
\end{mathletters}
and
\begin{mathletters}
\begin{eqnarray}
\ffive &\simeq&   \cnste \beta^{-{7\over2}} \fsfive
	   + \cnstf \eta^{3\over2} {5+3\eta\beta\over \sqrt{2+\eta\beta}}, \\
         \noalign{\medskip}
\ffiveth &=& \cnstf \eta^{3\over2} {5+3\eta\beta\over \sqrt{2+\eta\beta}},
\end{eqnarray}
\end{mathletters}
where
\begin{equation}
\fsone=\cases{ {1\over2} \left[ y A - \ln(y+A) \right]& if $y>0.05$\cr
          {1\over3} y^3-{1\over10} y^5+{3\over56} y^7-{5\over144} y^9+
         {35\over1408} y^{11}
                & if  $y\le 0.05$\cr},
\end{equation}
\begin{equation}
\fsthree=\cases {{1\over3} y^3 - {1\over2} \left[y A-\ln(y+A) \right]&
          if $y>0.05$\cr
    {1\over10} y^5 - {3\over56} y^7+{5\over144} y^9 - {35\over1408} y^{11}&
    if $y\le 0.05$\cr},
\end{equation}
and
\begin{equation}
\fsfive= \cases{{5\over8} y A \left(1 + {2\over5} y^2 \right) - {2\over3} y^3
		     - {5\over8} \ln(y+A)
         &if $y>0.1$\cr
       {1\over28} y^7 - {1\over36} y^9 + {15\over704} y^{11}
           &if $y\le 0.1$\cr}.
\end{equation}
The small $y$ expansions are used to avoid loss of accuracy due to strong
cancellations in nonrelativistic limit.

\subsection{Degenerate $\eta$-Derivatives}
The $\eta$ derivatives and the corresponding thermal functions are,
for all values of the relativity parameter $y$,
\begin{mathletters}
\begin{eqnarray}
\dfoneeta &\simeq& \cnste \eta^{1\over2}
(2+\eta\beta)^{1\over2}\left[1-\cnstg\,
{1\over\eta^2 (2+\eta\beta)^2}\right], \\
         \noalign{\medskip}
\dfoneetath &=&    -\cnstf\, {1\over\eta^{3\over2} (2+\eta\beta)^{3\over2}},
\end{eqnarray}
\end{mathletters}
\begin{mathletters}
\begin{eqnarray}
\dfthreeeta &\simeq& \cnste \eta^{3\over2}
(2+\eta\beta)^{1\over2}\left[1+\cnstg\,
{3+6\eta\beta+2\eta^2\beta^2\over\eta^2 (2+\eta\beta)^2}\right], \\
	 \noalign{\medskip}
\dfthreeetath &=&    \cnstf\, {3+6\eta\beta+2\eta^2\beta^2\over\eta^{1\over2}
(2+\eta\beta)^{3\over2}},
\end{eqnarray}
\end{mathletters}
\begin{mathletters}
\begin{eqnarray}
\dffiveeta &\simeq& \cnste \eta^{5\over2}
(2+\eta\beta)^{1\over2}\left[1+\cnstg\,
{15+20\eta\beta+6\eta^2\beta^2\over\eta^2 (2+\eta\beta)^2}\right], \\
	 \noalign{and}
\dffiveetath &=&  \cnstf \eta^{1\over2}\, {15+20\eta\beta+6\eta^2\beta^2\over
	 (2+\eta\beta)^{3\over2}}.
\end{eqnarray}
\end{mathletters}

\subsection{Degenerate $\beta$-Derivatives}
The $\beta$ derivatives and accompanying thermal terms are
\begin{mathletters}
\begin{eqnarray}
\dfonebeta &\simeq& \dfoneeta {\eta\over\beta} - {3\over2} \cnste
\beta^{-{5\over2}} \fsone + {1\over2} \cnstf \,
{1\over\eta^{1\over2}\beta}
{1+\eta\beta\over (2+\eta\beta) ^{1\over2}}, \\
             \noalign{\medskip}
\dfonebetath &=& \dfoneetath {\eta\over\beta} + {1\over2} \cnstf \,
{1\over\eta^{1\over2}\beta} {1+\eta\beta\over (2+\eta\beta) ^{1\over2}},
\end{eqnarray}
\end{mathletters}
\begin{mathletters}
\begin{eqnarray}
\dfthreebeta &\simeq& \dfthreeeta {\eta\over\beta} - {5\over2} \cnste
\beta^{-{7\over2}} \fsthree - {1\over2} \cnstf\, {\eta^{1\over2}
\over\beta} {3+2\eta\beta\over (2+\eta\beta) ^{1\over2}}, \\
	     \noalign{\medskip}
\dfthreebetath &=& \dfthreeetath {\eta\over\beta} - {1\over2} \cnstf \,
{\eta^{1\over2}\over\beta} {3+2\eta\beta\over (2+\eta\beta) ^{1\over2}},
\end{eqnarray}
\end{mathletters}
\begin{mathletters}
\begin{eqnarray}
\dffivebeta &\simeq& \dffiveeta {\eta\over\beta} - {7\over2} \cnste
\beta^{-{9\over2}} \fsfive  -{3\over2} \cnstf\,
{\eta^{3\over2}\over\beta}
{5+3\eta\beta\over (2+\eta\beta) ^{1\over2}}, \\
	     \noalign{and}
\dfthreebetath &=& \dffiveetath {\eta\over\beta} -{3\over2}   \cnstf \,
{\eta^{3\over2}\over\beta}
{5+3\eta\beta\over (2+\eta\beta) ^{1\over2}}.
\end{eqnarray}
\end{mathletters}

\section{Asymptotic Limits when Not Extremely Degenerate}
\subsection{Arbitrary Degeneracy}
In the relativistic ($\beta \gg 1)$ and nonrelativistic ($\beta \ll 1$)
limits the Fermi-Dirac functions  can be expressed
in terms of the simpler Fermi functions $G$, which only depend on $\eta$.
The Fermi integrals are given by
\begin{equation}
    G_k(\eta) = \int_{0}^{\infty} {{x^k}\over{\exp(x-\eta)+1}} \,dx
   \label{eq:fermii}
\end{equation}
with derivatives
\begin{equation}
    {\partial G_k\over \partial \eta} =
       \int_{0}^{\infty}{{x^k} \over
       {\left\{\exp\left[(x-\eta)\over 2\right]
       +\exp\left[(\eta-x)\over 2\right]\right\}^2}}\,dx.
\end{equation}
We require these functions for orders
$k=\slantfrac{1}{2}$, $1$, $\slantfrac{3}{2}$, $2$, $\slantfrac{5}{2}$,
$3$, and $\slantfrac{7}{2}$.
We have prepared, by numerical integration, tables of $G$ and
$\partial G/\partial\eta$, for each of those 7 orders, on a grid of
integral values of $-30 \le \eta \le 70$.
The evaluation of $G$ and $\partial G/\partial\eta$ in the following
formulae is accomplished by interpolation.
We discuss the tables in more detail below.

\subsubsection{Arbitrary Degeneracy and NonRelativistic}
In the nonrelativistic limit, $\beta < 10^{-6}$,
the Fermi-Dirac functions and their $\eta$-derivatives reduce to the
Fermi functions and their derivatives:
\begin{equation}
              F_k\fdarg \simeq G_k(\eta),
            \qquad k = \slantfrac{1}{2},\: \slantfrac{3}{2},\: \slantfrac{5}{2}
\end{equation}
and
\begin{equation}
      {\partial F_k\fdarg\over\partial\eta} \simeq {\partial
G_k(\eta)\over\partial\eta},
            \qquad k = \slantfrac{1}{2},\: \slantfrac{3}{2},\:
\slantfrac{5}{2}.
\end{equation}
The $\beta$-derivatives
\begin{equation}
      {\partial F_k\fdarg\over\partial\beta} \simeq  {1\over 4} G_{k+1}(\eta),
            \qquad k = \slantfrac{1}{2},\: \slantfrac{3}{2},\:
\slantfrac{5}{2}.
\end{equation}
involve the next higher order Fermi function.
The highest order derivative $\propto G_{7/2}$ only appears in the
expressions for the $\beta$-derivatives of $P$ and $E$,
multiplied by $\beta << 1$.  Because
$G_{7/2}$ is not otherwise required,  we carry the $G_{7/2}$ table in a
separate file for the convenience of implementations where setting
$\partial\ffive/\partial\beta = 0$ in the nonrelativistic limit is sufficient.

\subsubsection{Arbitrary Degeneracy and Extremely Relativistic}
In the relativistic limit, $\beta > 10^4$,
\begin{equation}
  F_k\fdarg =  2 \beta \left[{\partial F_k\fdarg\over\partial\beta}\right]
       \simeq  \sqrt{\beta\over2} \,G_{k+{1\over2}}(\eta),
            \qquad k = \slantfrac{1}{2},\: \slantfrac{3}{2},\: \slantfrac{5}{2}
\end{equation}
and
\begin{equation}
     {\partial F_k\fdarg\over\partial\eta}
     \simeq   \sqrt{\beta\over2} \, {\partial
G_{k+{1\over2}}(\eta)\over\partial\eta},
            \qquad k = \slantfrac{1}{2},\: \slantfrac{3}{2},\:
\slantfrac{5}{2}.
\end{equation}

\subsection{The NonDegenerate Limit --- $\eta < -30$}
In the nondegenerate limit, we make use of the well-known
\markcite{chandrasekhar}
(Chandrasekhar 1939) relations among the Fermi-Dirac integrals  and
the modified Bessel functions of the second  kind $\ki$ and $\kii$
(we shall henceforth not explicitly write the adjective ``modified'').
The Bessel functions are defined by
\begin{equation}
 \ki(\phi)= \int_{0}^{\infty} \cosh(t) \exp\left[-\phi \cosh(t)\right] dt
\end{equation}
    and
\begin{equation}
 \kii(\phi)= \int_{0}^{\infty} \cosh(2t) \exp\left[-\phi \cosh(t)\right] dt,
\end{equation}
 where
\begin{equation}
\cosh(x) = {\exp(x)+\exp(-x)\over2}
\end{equation}
is the hyperbolic cosine and
\begin{equation}\phi = {1\over \beta}.
\end{equation}
Defining
\begin{equation}
      {\cal B} = \cnste \exp\left(\eta + \phi\right),
          \label{eq:calb}
\end{equation}
the nondegenerate Fermi-Dirac functions are
\begin{mathletters}
\begin{equation}
  \fone \simeq {\cal B}{\phi}^{1\over2} \ki(\phi),
         \label{eq:bbon}
\end{equation}
\begin{equation}
  \fthree \simeq {\cal B}{\phi}^{3\over2} \left(\kii(\phi)-\ki(\phi)\right),
         \label{eq:bbto}
\end{equation}
\begin{equation}
  \ffive \simeq {\cal B} {\phi}^{5\over2}
            \left[2 \ki(\phi) + \left(3 \beta - 2\right) \kii(\phi)\right],
         \label{eq:bbth}
\end{equation}
\end{mathletters}
the Fermi-Dirac functions are the same as their
$\eta$-derivatives,
\begin{equation}
      {\partial F_k\fdarg\over\partial\eta} \simeq F_k\fdarg,
        \qquad k = \slantfrac{1}{2},\: \slantfrac{3}{2},\: \slantfrac{5}{2},
\end{equation}
and the $\beta$-derivatives are
\begin{mathletters}
\begin{equation}
  \dfonebeta \simeq {\cal B}\phi^{3\over2} \left[\phi\left(\kii-\ki\right)
                    - {3\over2}\ki\right],
         \label{eq:dbbon}
\end{equation}
\begin{equation}
  \dfthreebeta \simeq {\cal B}\phi^{5\over2} \left[2\phi\left(\ki - \kii\right)
                       + {1\over2}\left(5\ki + \kii\right)\right],
         \label{eq:dbbto}
\end{equation}
and
\begin{equation}
  \dffivebeta \simeq {\cal B}\phi^{5\over2}
      \left[{3\over2}\kii - 2\phi\left(2\ki + \kii\right)
      + 4\phi^2\left(\kii - \ki\right)\right].
         \label{eq:dbbth}
\end{equation}
\end{mathletters}

We make use of a combination of expansions and tabulations to evaluate
$\ki$ and $\kii$.

\subsubsection{Chebyshev Series Expansions}
Accurate Chebyshev series expansions exist for the Bessel functions.
\markcite{tooper} Tooper (1969) gives expansions, valid for $\phi > 8$,
for $K_0$ and $K_I$:
\begin{equation}
   K_j(\phi) = {e^{-\phi}\over \sqrt{\phi}} \,
          {\sum_{k=0}^\infty}^\prime\,a_k^j\, T_k\left({16\over\phi} - 1\right)
          \;\;j=0,I,
         \label{eq:tooper}
\end{equation}
where the $\prime$ in the summation means
\begin{equation}
          {\sum_{k=0}^\infty}^\prime\,a_k\, T_k \equiv
          {1\over2} a_0 + \sum_{k=1}^\infty\,a_k\, T_k.
\end{equation}
 (Tooper's expressions [Tables 13 and 14] for the Chebyshev expansions of
$K_0$ and $K_I$ contain an extra factor of
$\pi$.)
The Chebyshev polynomials $T_k(x)$ need not be explicitly calculated if the
series is
evaluated recursively.  Starting with
\begin{equation}
  b_{N+1}^j = b_{N+2}^j = 0,
  \label{eq:recura}
\end{equation}
successive iterates are given by
\begin{equation}
b_k^j = 2\left({16\over\phi} - 1\right) b_{k+1}^j
         - b_{k+2}^j + a_k^j,\;\;k = N, \cdots 0,
\end{equation}
and finally the sum
\begin{equation}
          {\sum_{k=0}^N}^\prime\,a_k^j\, T_k\left({16\over\phi} - 1\right)
            = {1\over2}\left(b_0^j - b_2^j\right).
\end{equation}
We use Tooper's coefficients from his Tables 13 ($j=0$) and 14 ($j=1$), for
which
$N=14$, to obtain
\begin{equation}
K_0(\phi) = {1\over2}{e^{-\phi}\over \sqrt{\phi}}\left(b_0^0 -
b_2^0\right),\;\;
K_I(\phi) = {1\over2}{e^{-\phi}\over \sqrt{\phi}}\left(b_0^I - b_2^I\right).
\end{equation}
The function $\kii$ is given by
\begin{equation}
  K_{II}(\phi) = K_0(\phi) + {2\over\phi} K_I(\phi).
             \label{eq:kii}
\end{equation}
For $\phi \le 8$, Chebyshev expansions from
\markcite{clenshaw} Clenshaw (1962) are applicable:
\begin{mathletters}
\begin{equation}
   K_0(\phi) = -\ln\left(\phi\over 8\right)\,
      {\sum_{k=0}^\infty}^\prime\,a_{2k}^{00}\, T_{2k}\left({\phi\over8}\right)
    + {\sum_{k=0}^\infty}^\prime\,a_{2k}^{01}\, T_{2k}\left({\phi\over8}\right)
\end{equation}
and
\begin{equation}
   K_I(\phi) = -\ln\left(\phi\over 8\right){\phi\over8}\,
      {\sum_{k=0}^\infty}^\prime\,a_{2k}^{I0}\, T_{2k}\left({\phi\over8}\right)
    + {1\over\phi} - {\phi\over8}\,
    {\sum_{k=0}^\infty}^\prime\,a_{2k}^{I1}\, T_{2k}\left({\phi\over8}\right)
    \label{eq:clenshaw}
\end{equation}
\end{mathletters}

Each of the  series in (\ref{eq:clenshaw}) is evaluated by
 recursion relations similar to those used in the evaluation of
(\ref{eq:tooper}).  Starting with (\ref{eq:recura}) and using
the Chebyshev coefficients $a_{2k}^{j}$ tabulated by Clenshaw for $N=17$,
the recursion
\begin{equation}
b_k^j = 2\left[2\left(\phi\over8\right)^2-1\right] b_{k+1}^j
          - b_{k+2}^j + a_{2k}^{j},\;\;k = N, \cdots 0,\;\;j = 00, 01, I0, I1
\end{equation}
is applied until the sum
\begin{equation}
   {\sum_{k=0}^N}^\prime\,a_{2k}^{j}\, T_{2k}(x)
   =  {1\over 2}(b_0^j - b_2^j), \;\;j = 00, 01, I0, I1.
\end{equation}

\subsubsection{NonDegenerate Small-$\beta$ Expansions}
In the nonrelativistic limit, $\beta \rightarrow 0$, a power series
asymptotic expansion (\markcite{gradshteyn} Gradshteyn \& Ryzhik 1980)
in $\beta$ is applicable.
After expanding $\ki$ and $\kii$ and collecting terms,
\begin{mathletters}
\begin{equation}
   \fone = \dfoneeta \simeq {\sqrt{\pi}\over2}\exp(\eta)
   \left(1 + {3\over 8}\beta  - {15\over 128}\beta^2
           + {105\over 1024}\beta^3 - {105\over 1024}\beta^4\right),
\end{equation}
\begin{equation}
     \fthree = \dfthreeeta \simeq  {3\sqrt{\pi}\over4}\exp(\eta)
      \left(1  + {5\over 8} \beta - {35\over 128} \beta^2
               - {2345\over 16384} \beta^3\right),
\end{equation}
\begin{equation}
     \ffive = \dffiveeta \simeq  {15\sqrt{\pi}\over8}\exp(\eta)
             \left(1 + {7\over 8} \beta - {539\over 4090} \beta^2\right),
\end{equation}
\end{mathletters}
\begin{mathletters}
\begin{equation}
\dfonebeta \simeq {3 \sqrt{\pi}\over 16}\exp(\eta)
     \left(1 - {5\over 8}\beta + {105\over 128}\beta^2
             - {35\over 32}\beta^3\right),
\end{equation}
\begin{equation}
 \dfthreebeta \simeq {15 \sqrt{\pi}\over 32}\exp(\eta)
      \left(1  - {7\over 8} \beta - {1407\over 2048} \beta^2\right),
\end{equation}
and
\begin{equation}
    \dffivebeta  \simeq {105 \sqrt{\pi}\over 34}\exp(\eta)
            \left(1 - {77\over 256} \beta\right).
  \label{eq:nddffb}
\end{equation}
\end{mathletters}
The higher order terms in the expansion (\ref{eq:nddffb}) have cancelled, but
since this expression is only used for $\beta < 10^{-3}$,
the remaining terms give sufficient accuracy.

\subsubsection{Extremely Relativistic and NonDegenerate}
In the relativistic limit, which we employ for $\beta > 10^4$,
$\ki(x) \simeq\ 1/x$, $\kii(x) \simeq 2/x^2$, and (\ref{eq:bbon})
through (\ref{eq:dbbth}) become
\begin{equation}
    F_k\fdarg = {\partial F_k\fdarg\over\partial\eta} \simeq
    \left(k+{1\over2}\right)! \, \exp\left(\eta+{1\over \beta}\right)
\sqrt{\beta\over2},
        \qquad k = \slantfrac{1}{2},\: \slantfrac{3}{2},\: \slantfrac{5}{2}
\end{equation}
and
\begin{equation}
      {\partial F_k\fdarg\over\partial\beta} \simeq
      \left({1\over2\beta} - {1\over\beta^2}\right) F_k\fdarg,
        \qquad k = \slantfrac{1}{2},\: \slantfrac{3}{2},\: \slantfrac{5}{2}.
\end{equation}

\subsubsection{Bessel Function Tabulation and Evaluation Methods}
We have prepared, by numerical integration, tabulations of
the Bessel functions on a grid of $-4 \le \log_{10}\phi \le 1.6$
with a spacing $\Delta\log_{10}\phi = 0.1$.

For $\phi > 1$, evaluations of derivatives by
(\ref{eq:dbbon}), (\ref{eq:dbbto}), and especially (\ref{eq:dbbth})
involve cancellation between the
$\ki$ and $\kii$ terms and require more precision than exists in
our table.
In addition,  $\ki$ and $\kii$
decrease ever more rapidly with increasing
$\phi$; this sharp falloff causes loss of  interpolation accuracy for
$\phi > 1$ ($\beta < 1$).  Greater accuracy is obtained by use of
high ($6^{\hbox{th}}$) order interpolation in $\ki$ and
$\kii$ and the second expression in equations \ref{eq:dbbon}, \ref{eq:dbbto},
and
\ref{eq:dbbth}, which  result in better cancellation, but the tables should
only
be used above $\beta = \beta_t$.  Between $\beta_t$ and the smaller $\beta_c$,
the
Chebyshev expansions should be used.  For $\beta < \beta_c$, the small-$\beta$
expansions are more accurate than the Chebyshev series, for which the accuracy
suffers as $\beta$ decreases.

For 6$^{\hbox{th}}$ order interpolation, the tables, rather than
the Chebyshev series are used for  $\beta > \beta_t = 10^{0.3}$.
Across this boundary, the Fermi-Dirac functions and derivatives match to
1 part in $5\times10^{5}$ or less.
For less accurate, lower order interpolations, larger values of $\beta_t$
may be appropriate (unless lower order interpolation is used for the sake of
reducing computer time).  The computer time taken to evaluate a set of
functions with the Chebyshev expansions is 1.09 times greater than evaluating
the same set with 6$^{\hbox{th}}$ order table interpolation.

When the $\beta$-derivatives are required, the small-$\beta$ expansions should
be used for $\beta < \beta_c = 10^{-2.8}$; for $\dffivebeta$, the
loss of accuracy with increasing or decreasing $\beta$ is steep away from
the respective side of the switching point.  When the
derivatives are not required,
$\log\beta_c = -2.0$ is recommended.  Across the $\beta_c = 10^{-2.8}$
boundary, the functions, the $\eta$-derivatives, and $\dfonebeta$ match to a
relative difference of $10^{-6}$ or better, $\dfthreebeta$ to $10^{-5}$, and
$\dffivebeta$ to 0.003.
The evaluation of the same set of functions
with the Chebyshev expansion requires 3.3 times as much computer time as with
the
small-$\beta$ expansions.

An alternative method of calculating
the nondegenerate quantities, which we do not employ, makes use the values of
the Fermi-Dirac functions and derivatives in the central table.  The $\eta$
dependence is given by equation \ref{eq:calb}, so that
$$ X\fdarg = \exp(\eta + 30)\,X(-30,\beta), $$
where $X$ is any $F_k$ or derivative.

\section{The Tables: Creation, Interpolation, and Accuracy}
\subsection{Numerical Integration}
All integrals considered here
were numerically evaluated using a 7-point adaptive
Newton--Cotes quadrature rule (implemented in the routine QNC79 from
the SLATEC software library).
For the Fermi-Dirac and Fermi integrals,
$\eta + 100$ (rather than $\infty$) was used for
the upper limit of the numerical integration; the lower limit
was $\max(0, \eta - 100)$,
whereas for the Bessel functions the integrations ran from
$0$ to $6 - \ln(\phi)$.
The integration routine evaluates each integral until a desired
accuracy is achieved.  The accuracy is expressed as a tolerance
${\cal E}$, where the result of the numerical integration $I$
does not deviate more than ${\cal E}I$ from the true answer.
We used ${\cal E} = 10^{-12}$.
A comparison of integrations made on 2 platforms---a Silicon Graphics
Indigo2 with an R4400 cpu chip running IRIX 5.3 and an Apollo DN4000
running DOMAIN/OS 10.3---agreed to within a relative difference of
$1.2\times 10^{-7}$.  This comparison suggests the accuracy of our
integrations is not better than 1 part in $10^{-7}$.
The values in the table are the Silicon Graphics integrations.

\subsection{The Tables}
Three tables of integrals accompany the electronic version
of this paper.
The file {\tt fdints.tab} contains the Fermi-Dirac integrals and the
Fermi integrals.
The file {\tt dfdints.tab} contains the derivatives of the Fermi-Dirac and
of the Fermi integrals.
The Bessel functions are found in {\tt bessel.tab}.
The file {\tt fermi7h.tab} contains the Fermi integrals of order
$\slantfrac{7}{2}$.
These tables are read by the FORTRAN subroutine {\tt calcdfi},
which is also provided
along with the tables; the exact format of the tables can be found
by examining the relevant {\tt READ} statements in the subroutine.
The functions and the derivatives are maintained as separate files to
facilitate implementations where only the function values are required;
subroutine {\tt calcfi}, also supplied, is a variant of {\tt calcdfi}
with the derivatives stripped out.

Subroutine {\tt calcdfi} reads the data from the binary (or unformatted) files
{\tt fdints.unf} (which contains the Bessel data)
and {\tt dfdints.f} is they both exist; if they do not exist,
the formatted files {\tt fdints.tab}, $\cdots$ are read, and the logarithm of
the
data are written to the binary files.
The binary files are preferred because the formatted files
are 5 times as large and take 35 times longer to load.
All data is held in memory as (natural) logarithms.

The main table of the Fermi-Dirac functions and their integrals for
             $-30 \le \eta \le 70$ and $-6 \le \log\beta \le 4$,
the same ranges as used in the tabulations in Appendix A.2 of
\markcite{cox} Cox \&\ Giuli (1968).
The table uses a finer grid for small values of $\vert\eta\vert$
and $\vert\beta\vert$.
The table contains 21 $\beta$ values corresponding to the integral and
half-integral values of $-6 \le \log(\beta) \le 4$.
The 47 points in the $\eta$ grid are concentrated towards
$\eta = 0$.  The spacing is 0.1 for $-1 \le \eta \le 1$,
1 for $-5 \le \eta \le -1$ and $1 \le \eta \le 5$,
and 5 for  $-30 \le \eta \le -5$  and $5 \le \eta \le 70$ .
The $\eta$ grid for the Fermi integrals and derivatives consists of 101 values
of integral $-30 \le \eta \le 70$.
The $\phi$ grid for the Bessel functions contains 59 values
with a spacing $\Delta\log_{10}\phi = 0.1$.

\subsection{Interpolation}
All interpolation is made with the logarithms of the integrals and
derivatives as functions of $\log\beta$ (or $\log\phi$) and $\eta$
(except as noted in the discussion of second order interpolation
 in the 2-dimensional table).  Use of the direct values, rather than the
logarithms, is much less accurate for all interpolation orders.

\subsubsection{Second Order Interpolation}
Second order interpolation for the 1-dimensional functions is a simple
linear interpolation between the values on bracketing grid points.
For the central 2-dimensional tables, the interpolation
is logarithmic in the $\eta$ direction
\begin{equation}
      f(\beta) = \exp\left\{ (1-d) \ln[f(\eta_a,\beta)]
                              + d  \ln[f(\eta_b,\beta)]\right\}
\end{equation}
where $d$ is the interpolation coefficient and $\eta_{(a,b)}$
are points in the table.
The ${2\over3}$ power is used for the interpolation in the $\beta$ direction,
\begin{equation}
   f = \left[ (1-d) f(\beta_a)^{2\over3}
               + d  f(\beta_b)^{2\over3}\right]^{3\over2}
\end{equation}
where $d$ and $\beta_{(a,b)}$ have meanings similar to the above.

\subsubsection{Higher Order Polynomial Interpolation}
Our polynomial interpolation is based on the subroutines {\tt POLINT} and
{\tt POLIN2} from the {\it Numerical Recipes} book
(\markcite{press} Press, et al. 1986).
We examined the accuracy and speed of the interpolation for several polynomial
orders.  We found $6^{\rm th}$ order to be most accurate, but with a
substantial
penalty in execution speed.  Orders 2 through 6 are available in the subroutine
we provide.

The tables below give the accuracy and execution time for various orders and
functions.  The accuracy is given in terms of the relative error
$\max(f/f\prime, f\prime/f)$, where $f$ is truth, as given by a numerical
integration, and $f\prime$ is the value from interpolation.
Execution times are normalized to 1 for $2^{\rm nd}$ order.  The timings were
made with uniform samplings over the respective tables.

\subsubsection{The Two Dimensional Tables}
Table~\ref{tbl-1} gives the maximum relative error for the Fermi-Dirac
integrals,
along with the execution times, and Table~\ref{tbl-2} gives the maximum
relative
errors for the derivatives.  The accuracy is based on comparisons with
numerical integrations made at $\eta = 0.6\eta_i + 0.4\eta_{i+1}$ and
$\log\beta = 0.6\log\beta_j + 0.4\log\beta_{j+1}$ for each $(i,j)$ cell
in the table, a similar set of integrations with $0.4 \leftrightarrow 0.6$
for every other $(i,j)$ cell, and a sampling of cells with a set of
10 integrations crossing the cell in some direction.

The order in which the 1-dimensional interpolations were performed
(ie. $\eta$ direction first or $\beta$ direction first)
made no difference in the accuracy.

\placetable{tbl-1}
\placetable{tbl-2}

\subsubsection{Fermi Integrals}
Table~\ref{tbl-3} gives the maximum error in the Fermi integrals and
their derivatives for a range of $\pm 0.5$ in $\eta$ centered on
the value listed.  There is variation of up to 100\%
among the relative errors for the individual functions and derivatives.
The maximum relative errors for the 6 functions and the 6 derivatives
are the same to within 1\%.
The relative execution times for the interpolations are given in
Table~\ref{tbl-4}.

\placetable{tbl-3}
\placetable{tbl-4}

\subsubsection{Bessel Function Interpolation}
Table~\ref{tbl-5} lists the maximum relative errors in the
Bessel functions for several ranges of $\phi$.
For each $\phi$ range in each order, the maximum relative errors for
$\ki$ and $\kii$ were comparable.
The larger of the two is listed in the table.
The error becomes increasingly worse as $\phi$ increases and the Bessel
functions
are dominated by the factor $\exp{-\phi}$.

\placetable{tbl-5}

\subsubsection{Accuracy at Treatment Boundaries}
Different methods are used to evaluate the Fermi-Dirac integrals and their
derivatives for different ranges of $\eta$ and $\beta$.  The closeness of the
evaluations on opposite sides of a treatment boundary is given in
Table~\ref{tbl-6}.
The quantity tabulated is the largest value of
$\max(f_r/f_l,f_l/f_r)$ along a boundary.  For an $\eta$ boundary,
$f_{(r,l)} = f(\eta\pm 0.000001)$, and for a $\beta$ boundary
$f_{(r,l)} = f(\log\beta\pm 0.000001)$.
The results are most excellent, especially for
$6^{\rm th}$ order interpolation.

The $\beta$-derivatives lose all accuracy near $\log\beta = -1.6$.
Accordingly, we
also list the closeness along $\eta = -30$ excluding a range near $\log\beta =
-1.6$.

\placetable{tbl-6}

\acknowledgments

We are grateful to Onno Pols for helping us correctly implement the
Eggleton, Faulkner, \&\ Flannery formulae, and to Jim Lattimer for
discussions on various approximations.

\clearpage

\begin{deluxetable}{crccc}
\tablecaption{Main Table Interpolation. \label{tbl-1}}
\tablewidth{0pc}
\tablehead{
   \colhead{}& \colhead{}   &\multicolumn{3}{c}{Maximum Relative Error}\\
   \cline{3-5}\\
 \colhead{N}  & \colhead{time} &
   \colhead{$\fone$} & \colhead{$\fthree$} & \colhead{$\ffive$}
}
\startdata
 2  &  1.0      & 1.082 & 1.118 & 1.143 \nl
 3  &  1.8      & 1.572 & 1.646 & 1.628 \nl
 4  &  2.9      & 1.023 & 1.023 & 1.021 \nl
 5  &  4.6      & 1.027 & 1.016 & 1.009 \nl
 6  &  7.1      & 1.007 & 1.003 & 1.004 \nl
 7  & 10.\phn   & 1.005 & 1.010 & 1.012 \nl
 8  & 14.\phn   & 1.013 & 1.007 & 1.005 \nl
 9  & 19.\phn   & 1.096 & 1.050 & 1.026 \nl
\enddata
\end{deluxetable}
\clearpage

\begin{deluxetable}{ccccccc}
\tablecaption{Main Table Derivative Interpolation. \label{tbl-2}}
\tablewidth{0pc}
\tablehead{
  \colhead{}   &\multicolumn{6}{c}{Maximum Relative Error}\\
   \cline{2-7}\\
 \colhead{N}  & \colhead{$\dfoneeta$} & \colhead{$\dfthreeeta$} &
\colhead{$\dffiveeta$}
  & \colhead{$\dfonebeta$} & \colhead{$\dfthreebeta$} & \colhead{$\dffivebeta$}
}
\startdata
 2   & 1.034 & 1.082 & 1.118 & 1.132 & 1.159 & 1.177\nl
 3   & 1.391 & 1.572 & 1.646 & 1.573 & 1.646 & 1.628\nl
 4   & 1.019 & 1.023 & 1.024 & 1.024 & 1.023 & 1.018\nl
 5   & 1.032 & 1.027 & 1.016 & 1.022 & 1.010 & 1.010\nl
 6   & 1.011 & 1.007 & 1.003 & 1.005 & 1.004 & 1.005\nl
 7   & 1.019 & 1.006 & 1.010 & 1.007 & 1.012 & 1.013\nl
 8   & 1.026 & 1.013 & 1.007 & 1.010 & 1.006 & 1.005\nl
 9   & 1.187 & 1.096 & 1.050 & 1.069 & 1.036 & 1.019\nl
\enddata
\end{deluxetable}
\clearpage
\begin{deluxetable}{rccccc}
\tablecaption{Fermi Function Interpolation Relative Errors. \label{tbl-3}}
\small
\tablewidth{0pc}
\tablehead{
  \colhead{}   &\multicolumn{5}{c}{Interpolation Order}\\
   \cline{2-6}\\
 \colhead{$\eta$\phn\phn} & \colhead{2} & \colhead{3} &
                            \colhead{4} & \colhead{5} & \colhead{6}
}
\startdata
 $-29.5$ & 1.0000006 & 1.0000006 & 1.0000006 & 1.0000006 & 1.0000006 \nl
 $-19.5$ & 1.0000006 & 1.0000006 & 1.0000006 & 1.0000006 & 1.0000006 \nl
 $-14.5$ & 1.0000005 & 1.0000023 & 1.0000007 & 1.0000008 & 1.0000006 \nl
 $ -9.5$ & 1.0000068 & 1.0000182 & 1.0000011 & 1.0000006 & 1.0000005 \nl
 $ -4.5$ & 1.0009817 & 1.0025651 & 1.0001572 & 1.0000744 & 1.0000216 \nl
 $ -2.5$ & 1.0063697 & 1.0168104 & 1.0008456 & 1.0003333 & 1.0000158 \nl
 $ -1.5$ & 1.0137371 & 1.0364101 & 1.0011945 & 1.0002281 & 1.0002464 \nl
 $ -0.5$ & 1.0225276 & 1.0566500 & 1.0005099 & 1.0009853 & 1.0002920 \nl
   0.5 & 1.0247931 & 1.0412589 & 1.0026541 & 1.0018509 & 1.0004923 \nl
   1.5 & 1.0187399 & 1.0212393 & 1.0022867 & 1.0005230 & 1.0003849 \nl
   2.5 & 1.0132760 & 1.0471996 & 1.0007911 & 1.0016337 & 1.0002560 \nl
   4.5 & 1.0091014 & 1.0247539 & 1.0006357 & 1.0002153 & 1.0000434 \nl
   9.5 & 1.0041356 & 1.0051802 & 1.0000569 & 1.0000178 & 1.0000018 \nl
  14.5 & 1.0020823 & 1.0019885 & 1.0000169 & 1.0000033 & 1.0000004 \nl
  19.5 & 1.0012197 & 1.0008866 & 1.0000060 & 1.0000016 & 1.0000005 \nl
  24.5 & 1.0007939 & 1.0004615 & 1.0000026 & 1.0000006 & 1.0000005 \nl
  29.5 & 1.0005556 & 1.0002668 & 1.0000014 & 1.0000004 & 1.0000004 \nl
  34.5 & 1.0004097 & 1.0001662 & 1.0000010 & 1.0000005 & 1.0000005 \nl
  39.5 & 1.0003149 & 1.0001134 & 1.0000009 & 1.0000006 & 1.0000005 \nl
  44.5 & 1.0002490 & 1.0000782 & 1.0000005 & 1.0000005 & 1.0000005 \nl
  49.5 & 1.0002017 & 1.0000569 & 1.0000004 & 1.0000007 & 1.0000005 \nl
  59.5 & 1.0001400 & 1.0000308 & 1.0000005 & 1.0000006 & 1.0000005 \nl
  69.5 & 1.0001032 & 1.0000216 & 1.0000021 & 1.0000034 & 1.0000067 \nl
\enddata
\end{deluxetable}
\clearpage

\begin{deluxetable}{crr}
\tablecaption{Interpolation Times. \label{tbl-4}}
\tablewidth{0pc}
\tablehead{
  \colhead{} &\multicolumn{2}{c}{Relative Time}\\
  \cline{2-3}\\
 \colhead{N} &\multicolumn{1}{c}{$G$, $\partial G/\partial\eta$} &
              \multicolumn{1}{c}{$\ki$, $\kii$}
}
\startdata
2 &  1.0      &  1.0 \nl
3 &  7.6      &  4.9 \nl
4 & 11.\phn   &  7.1 \nl
5 & 16.\phn   &  9.8 \nl
6 & 22.\phn   & 13.\phn  \nl
\enddata
\end{deluxetable}
\clearpage

\begin{deluxetable}{ccccccc}
\tablecaption{Bessel Function Interpolation Relative Errors. \label{tbl-5}}
\tablewidth{0pc}
\tablehead{
  \colhead{}   &\multicolumn{5}{c}{Interpolation Order}\\
   \cline{3-7}\\
 \colhead{$\log\phi_a$} & \colhead{$\log\phi_b$} & \colhead{2} &
               \colhead{3} & \colhead{4} & \colhead{5} & \colhead{6}
}
\startdata
    $-4.0$ &      $-2.0$  &  1.000066  &  1.000139 &  1.000042  &  1.000047 &
1.000042 \nl
    $-2.0$ &  \phm{$-$} 0.0  &  1.005176  &  1.008385 &  1.000117  &  1.000050
&   1.000041 \nl
\phm{$-$} 0.0 &  \phm{$-$} 0.5  &  1.018382  &  1.028126 &  1.000306  &
1.000092 &   1.000059 \nl
\phm{$-$} 0.5 &  \phm{$-$} 1.0  &  1.060710  &  1.090631 &  1.000918  &
1.000201 &   1.000088 \nl
\phm{$-$} 1.0 &  \phm{$-$} 1.5  &  1.205830  &  1.311856 &  1.002875  &
1.000673 &   1.000389 \nl
\enddata
\end{deluxetable}
\clearpage

\begin{deluxetable}{lcccccc}
\tablecaption{Difference at Treatment Boundaries. \label{tbl-6}}
\tablewidth{0pc}
\tablehead{
  \colhead{}   &\multicolumn{5}{c}{Interpolation Order}\\
   \cline{3-7}\\
 \colhead{Boundary} & \colhead{Function} & \colhead{2} &
               \colhead{3} & \colhead{4} & \colhead{5} & \colhead{6}
}
\startdata
$\eta = 70$ & $F$                        & 1.0007 & 1.0026 & 1.0022 & 1.0582 &
1.0309  \nl
$\eta = 70$ & $\partial F/\partial\eta$  & 1.0008 & 1.0024 & 1.0030 & 1.0632 &
1.0308  \nl
$\eta = 70$ & $\partial F/\partial\beta$ & 1.0011 & 1.0023 & 1.0038 & 1.0611 &
1.0279  \nl
\noalign{\smallskip}
$\eta = -30$ & $F$                        & 1.0027\tablenotemark{a} & 1.0022 &
1.0029 & 1.0473 & 1.0303 \nl
$\eta = -30$ & $\partial F/\partial\eta$  & 1.0027\tablenotemark{a} & 1.0022 &
1.0029 & 1.0473 & 1.0303 \nl
$\eta = -30$ & $\partial F/\partial\beta$ & 1.0021\tablenotemark{a} & 1.0020 &
1.0024 & 1.0445 & 1.0279 \nl
\noalign{\smallskip}
$\log \beta = 4$ & $F$                        & 1.0072 & 1.0221 & 1.0231 &
1.5966 & 1.1481  \nl
$\log \beta = 4$ & $\partial F/\partial\eta$  & 1.0081 & 1.0301 & 1.0231 &
1.5967 & 1.1270  \nl
$\log \beta = 4$ & $\partial F/\partial\beta$ & 1.0064 & 1.0207 & 1.0225 &
1.5971 & 1.1481  \nl
\noalign{\smallskip}
$\log \beta = -6$ & $F$                        & 1.0071 & 1.0262 & 1.0227 &
1.5976 & 1.1386  \nl
$\log \beta = -6$ & $\partial F/\partial\eta$  & 1.0101 & 1.0312 & 1.0227 &
1.5752 & 1.1132  \nl
$\log \beta = -6$ & $\partial F/\partial\beta$ & 1.0042 & 1.0134 & 1.0213 &
1.5864 & 1.1258  \nl
\enddata
\tablenotetext{a}{The difference is 1.0007 when the range
                        $3.7 \le \log\beta \le 4.0$ is excluded}
\end{deluxetable}

\end{document}